\documentclass[conference]{IEEEtran}
\IEEEoverridecommandlockouts
\usepackage{cite}
\usepackage{amssymb,amsfonts}
\usepackage{algorithmic}
\usepackage{graphicx}
\usepackage{textcomp}
\usepackage{xcolor}
\usepackage{subfig}
\usepackage{amsmath}
\usepackage{balance}
\usepackage[english]{babel}
\usepackage{blindtext}
\usepackage{array}
\usepackage{nicefrac}
\def\BibTeX{{\rm B\kern-.05em{\sc i\kern-.025em b}\kern-.08em
    T\kern-.1667em\lower.7ex\hbox{E}\kern-.125emX}}
\begin{document}

\title{Transaction Placement in Sharded Blockchains \\
\thanks{This research is supported in part by a Ripple Graduate Fellowship.}
}

\author{\IEEEauthorblockN{1\textsuperscript{st} Liuyang Ren}
\IEEEauthorblockA{
\textit{University of Waterloo}\\
Waterloo, Canada\\
l27ren@uwaterloo.ca}
\and
\IEEEauthorblockN{2\textsuperscript{nd} Paul A. S. Ward}
\IEEEauthorblockA{
\textit{University of Waterloo}\\
Waterloo, Canada\\
pasward@uwaterloo.ca}
}

\maketitle

\begin{abstract}
While many researchers adopt a sharding approach to design scaling blockchains, few works have studied the transaction placement problem incurred by sharding protocols. The widely-used hashing placement algorithm renders an overwhelming portion of transactions as cross-shard. 
In this paper, we analyze the high cost of cross-shard transactions and reveal that most Bitcoin transactions have simple dependencies and can become single-shard under a placement algorithm taking transaction dependencies into account.
In addition, we perform a case study of OptChain,  which is the state-of-the-art transaction placement algorithm for sharded blockchains, and find a shortcoming of it.
A simple fix is proposed, and our evaluation results demonstrate that the proposed fix effectively helps OptChain overcome the shortcoming and significantly improve the system performance under a special workload. The authors of OptChain made some revisions to the algorithm description after noticing our work. Their updated algorithm does not exhibit the shortcoming under the workloads employed by this paper.
\end{abstract}


\section{Introduction}
\label{sec:intro}
Conventional blockchain systems suffer from low performance due to the PoW consensus protocol, which requires all nodes to duplicate the works from each other. To improve the performance, various designs have been proposed---e.g., shortening block intervals \cite{croman2016scaling}, incorporating off-chain blocks \cite{lewenberg2015inclusive}\cite{li2018scaling}, allowing one miner to consecutively propose multiple blocks \cite{eyal2016bitcoin}, sharding \cite{luu2016secure_elastico}\cite{kokoris2018omniledger}\cite{zamani2018rapidchain}\cite{dang2019towards}, journaling aggregated transaction effects to blockchains \cite{decker2015fast} \cite{poon2016bitcoin}, etc.
Among these techniques, sharding is a promising approach and has been explored by many researchers. The high-level idea of sharding is to partition a system into multiple shards and distribute its workloads to different shards for parallel processing so that the system performance scales with the number of nodes.
However, because each shard usually stores a disjoint subset of the system state \cite{kokoris2018omniledger}\cite{zamani2018rapidchain}\cite{dang2019towards}, transactions modifying more than one such subsets would inevitably incur cross-shard communication. Specifically, such cross-shard transactions rely on Atomic Commit Protocols (ACPs) to guarantee state consistency between shards. In addition to network bandwidth consumption, ACPs also put extra computational burdens on the involved shards, especially in trustless environments (i.e., public blockchains), where expensive digital signatures are employed to prove message authenticity. 

\begin{figure}[htbp]
  \centerline{\includegraphics[width=0.7\linewidth]{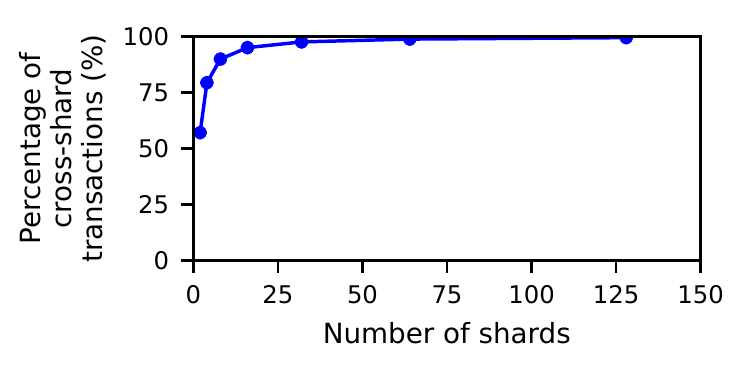}}
  \caption{A high percentage of transactions are cross-shard under the hashing placement algorithm.  }
   \label{fig:cross_shard_percent_intro}
\end{figure}

While many previous works focus on blockchain sharding protocols, few researchers have paid attention to transaction placement, which is about selecting a shard to process a given transaction. Transaction placement algorithms have a major impact on the number of cross-shard transactions and consequently the system performance.
The transaction placement algorithm used by most, if not all, sharding protocols is the Hashing Placement (HP) \cite{luu2016secure_elastico}. HP places transactions to shards based on the prefix matching of transaction IDs and shard IDs. Because transaction IDs are hash values (e.g., SHA256 values in Bitcoin), and hash values are uniformly distributed over the output range of the corresponding hash function \cite{stinson2005cryptography}, HP is equivalent to placing transactions randomly. As a result, HP produces a considerable number of cross-shard transactions. Fig.~\ref{fig:cross_shard_percent_intro} illustrates the placement outcome for the 1.57 million non-coinbase\footnote{Coinbase transactions have no input UTXOs and credit newly ``minted" UTXOs to miners. Non-coinbase transactions spend existing UTXOs.} transactions in the first 150k Bitcoin blocks. When the system involves more than 32 shards, almost all non-coinbase transactions are cross-shard.

This work has three contributions: 1) revealing that the overwhelming majority of Bitcoin transactions have only one parent transaction; 2) identifying the root cause of the high cost of cross-shard transactions; and 3) identifying a shortcoming in OptChain \cite{nguyen2019optchain}, a state-of-the-art blockchain transaction placement algorithm, and proposing an effective fix.

\section{Background}
A blockchain is an ordered set of blocks, each of which contains an ordered set of transactions. Every node maintains a local copy of the whole blockchain, executes the transactions in the order specified by the blockchain, and thus ends up with the same system state as other nodes.

\subsection{Unspent Transaction Output (UTXO) Model}

\begin{figure}[htbp]
  \centerline{\includegraphics[trim={0 0cm 0cm 0cm},width=0.65\linewidth]{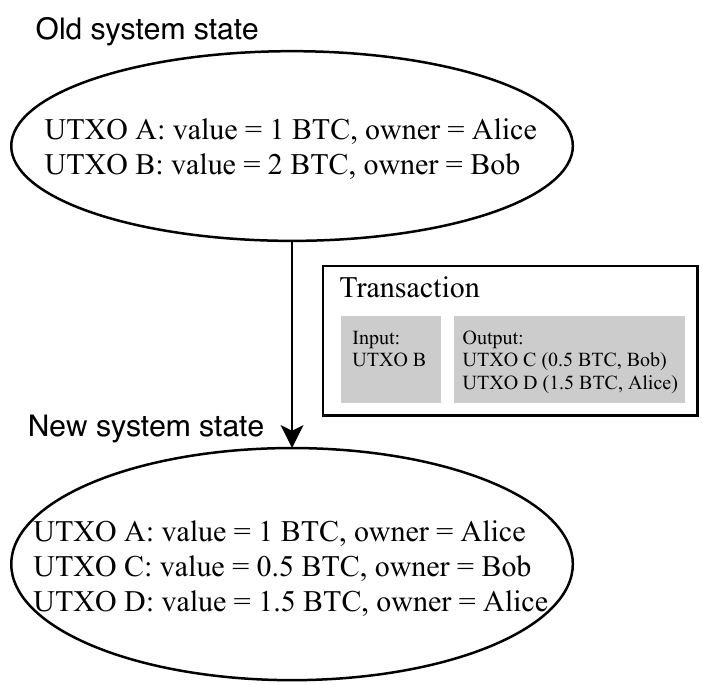}}
\caption{Transaction execution}
  \label{fig:tx_exe}
\end{figure}

Unlike conventional banking systems, cryptocurrencies like Bitcoin use a UTXO model to express their system states instead of the account-balance model. Accordingly, a transaction spends input UTXOs and generates output UTXOs. Figure \ref{fig:tx_exe} demonstrates the transaction execution in Bitcoin. Bob sends 1.5 BTC to Alice by creating a transaction that spends his 2-BTC UTXO and generates a 1.5-BTC UTXO for Alice as well as a 0.5-BTC UTXO for himself. Once the transaction gets executed, the 2-BTC UTXO (i.e., \emph{UTXO B} in Fig.~\ref{fig:tx_exe}) does not exist anymore.  All transactions have at least one input UTXO, except for coinbase transactions, which spend nothing and credit output UTXOs to miners. 

The transaction placement problem in a UTXO-based system is to find a shard to process a transaction  and store the output UTXOs of the transaction. This shard is called the \emph{output shard} of this transaction. Similarly, a shard storing at least one input UTXO of this transaction is called an \textit{input shard}.

\subsection{Blockchain Sharding Protocols}
In this section, we focus on describing the sharding protocol OmniLedger \cite{kokoris2018omniledger} and briefly introduce other sharding protocols because OmniLedger is used by OptChain for evaluation. 

\begin{figure*}[hbtp]
\vspace{1cm}
\centering
\subfloat[A cross-shard transaction]{\includegraphics[trim={-0.8cm -0.2cm -0.8cm 0.8cm}, width=0.2\linewidth]{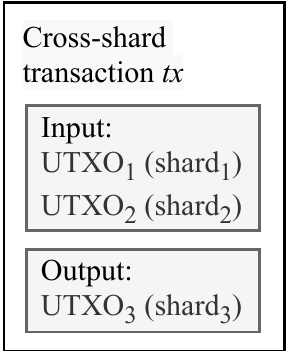}%
\label{fig:atomix_tx}}
\hspace{0.2cm}
\subfloat[Lock-success scenario in Atomix]{\includegraphics[trim={0 0cm 0cm 0.8cm}, width=0.62\linewidth]{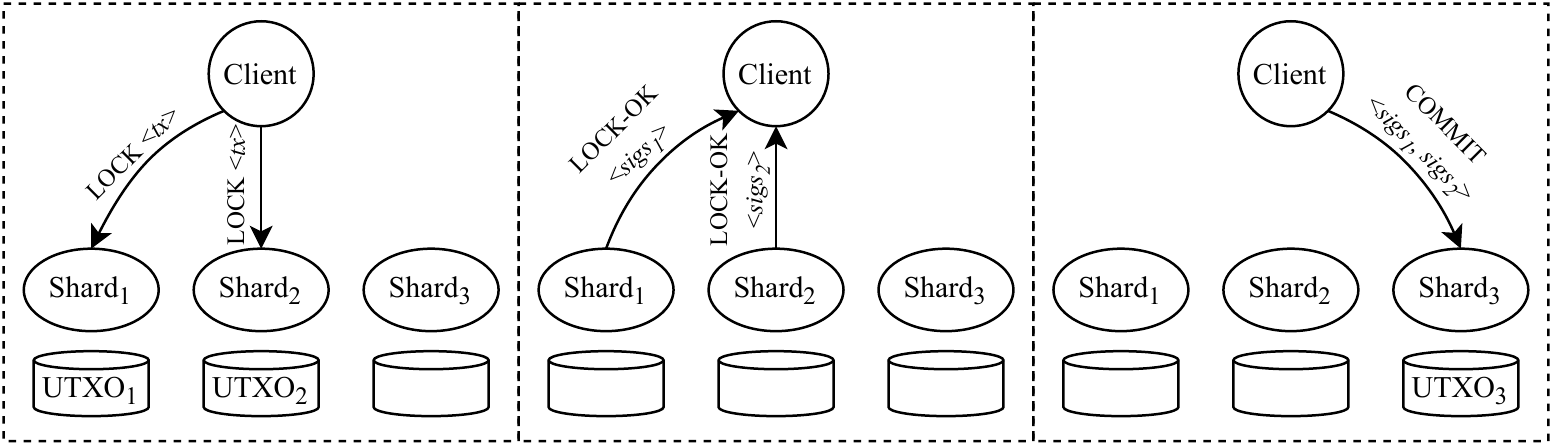}%
\label{fig:atomix_processing}} 
\caption{An example of  OmniLedger's Atomix protocol}
\label{fig:atomix}
\end{figure*}

Blockchain sharding protocols divide nodes into shards that process transactions in parallel. 
OmniLedger shards maintain their own respective ledgers. Every shard produces blocks extending from its individual ledger, and the blockchain is the union of the ledgers of all shards.
Shard members are selected using Proof-of-Work (PoW) to protect the system from Sybil attacks and periodically reconfigured so that attackers cannot corrupt a shard by attacking a small and fixed group of nodes. 
Within a shard, nodes run the Practical Byzantine Fault Tolerance (PBFT) protocol \cite{castro1999practical} to agree on the validity and the execution order of transactions. 

Cross-shard transactions rely on atomic commit protocols (ACPs) to be either unanimously committed or unanimously aborted. The problem of guaranteeing transaction atomicity dates back to the late 1970s \cite{lampson1979crash}\cite{gray1978notes}. Among various ACPs, the two-phase commit protocol  (2PC) \cite{lampson1979crash} is the most widely used one in the industry and academia \cite{guerraoui2017fast}\cite{skeen1981nonblocking}. OmniLedger has proposed a similar protocol called \emph{Atomix} for atomically processing transactions across shards. The basic idea of Atomix is shown in Fig. \ref{fig:atomix}. The client requests the input shards to lock the input UTXOs of its cross-shard transaction, and the input shards reply with signed lock results. 
If all input shards reply with positive lock results, the client then sends the output shard a \texttt{COMMIT} request along with signatures from the input shards as proofs of successful locking. 
The output shard adds the output UTXO to its system state, provided all signatures of the input shards are valid. If any input shard cannot lock an input UTXO, it informs the client about the failure with a signed message, and the client will use the message as a proof to request other input shards to unlock their respective input UTXOs.
In blockchain environments, it is challenging to find an ACP coordinator, whose misbehavior may lead to forever-locked UTXOs. OmniLedger chooses clients as the coordinators of their own transactions so that coordinators are incentivized to conform to the ACP protocol.

Like OmniLedger, Elastico \cite{luu2016secure_elastico} and AHL \cite{dang2019towards} also employ PBFT as their intra-shard consensus protocol.  
RapidChain \cite{zamani2018rapidchain}, on the other hand, uses a variant of the synchronous consensus protocol proposed by Ren et al. \cite{ren2017practical} to tolerate the same number of faulty nodes with a smaller shard size than PBFT. 
In terms of ACPs, Dang et al. \cite{dang2019towards} incorporates 2PC and leverages an entire Byzantine fault-tolerant committee as the coordinator. 
RapidChain [47] processes a cross-shard transaction by splitting it into multiple sub-transactions, each of which spends UTXOs belonging to the same shard (i.e., every sub-transaction has only one input shard). 
Elastico is fundamentally different from other sharding protocols in that it does not partition system states despite the fact that transactions are partitioned. In other words, every Elastico node still maintains the whole system state (i.e., all UTXOs). As a result, there are no cross-shard transactions in Elastico.
Table \ref{tab:cross_protocol} summarizes these sharding protocols.

\begin{table*} [htbp]
\centering
  \caption{Comparison of Various Blockchain Shading Protocols}
  \label{tab:cross_protocol}
  \begin{tabular}{p{0.075\textwidth}|p{0.297\textwidth}|p{0.157\textwidth}|p{0.18\textwidth}|p{0.142\textwidth}}
    \hline
    Protocol & Intra-shard Consensus & ACP & ACP Coordinator & Transaction Placement\\
    \hline
    Elastico & PBFT & N/A & N/A & Hashing placement\\
    \hline
    OmniLedger & PBFT & Atomix \cite{kokoris2018omniledger} & Client & Hashing placement \\
    \hline
    RapidChain & Practical synchronous byzantine consensus \cite{ren2017practical} & Transaction 
    Splitting \cite{zamani2018rapidchain} & PBFT leader of output shard & Hashing placement \\
    \hline
    AHL & PBFT & 2PC & Dedicated BFT committee  & Hashing placement\\
   \hline
\end{tabular}
\end{table*}

\section{Transaction Characteristics}
Understanding transaction characteristics allows us to determine whether carefully placing transactions to shards can achieve substantial performance improvement. We are particularly interested in two characteristics: 1) how many transactions can be easily turned into single-shard transactions, and 2) how expensive a cross-shard transaction is when compared with a single-shard transaction. A transaction placement algorithm can only benefit the system performance if cross-shard transactions cost a lot more than their single-shard counterparts and can be greatly reduced by the algorithm. 

\subsection{Transaction Dependencies}
If transaction $tx2$ spends the UTXO(s) produced by transaction $tx1$, then $tx1$ is called the parent transaction and $tx2$ is the child transaction.  
Given a transaction, the number of its parent transactions directly affects its probability of being single-shard. For example, a transaction with only one parent can be a single-shard transaction if it is placed to the same shard as its parent. In contrast, a transaction with 100 parents is very unlikely to be single-shard since that requires all of the 100 parents have been placed to the same shard. 
Thus, we analyze parent counts for the 1,119,794 transactions in the 601001st to 601500th Bitcoin block  (coinbase transactions are excluded since they have no parents and are guaranteed to be single-shard). 
Because Bitcoin blocks are usually referred to using \emph{height}s, which are indices starting from zero, we refer to the $(k+1)$-th block as block $k$ in the rest of this paper.

\begin{figure}[htbp]
  \centerline{\includegraphics[trim={0cm 0.5cm 0cm 0.4cm},width=0.75\linewidth]{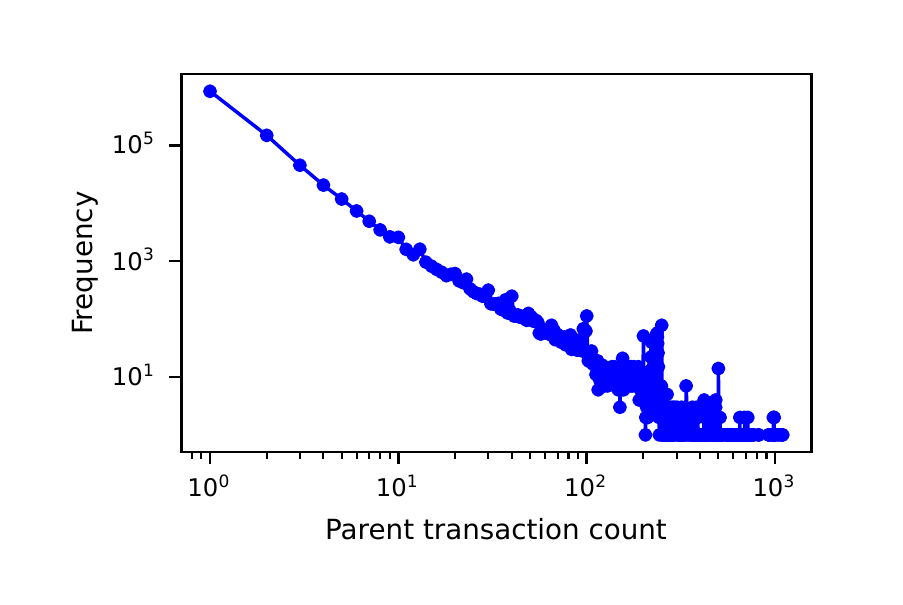}}
  \caption{Distribution of parent transaction counts
  }
   \label{fig:input_tx_cnt}
\end{figure}


\begin{figure}[htbp]
  \centerline{\includegraphics[trim={0cm 0.3cm 0cm 0.5cm},width=0.75\linewidth]{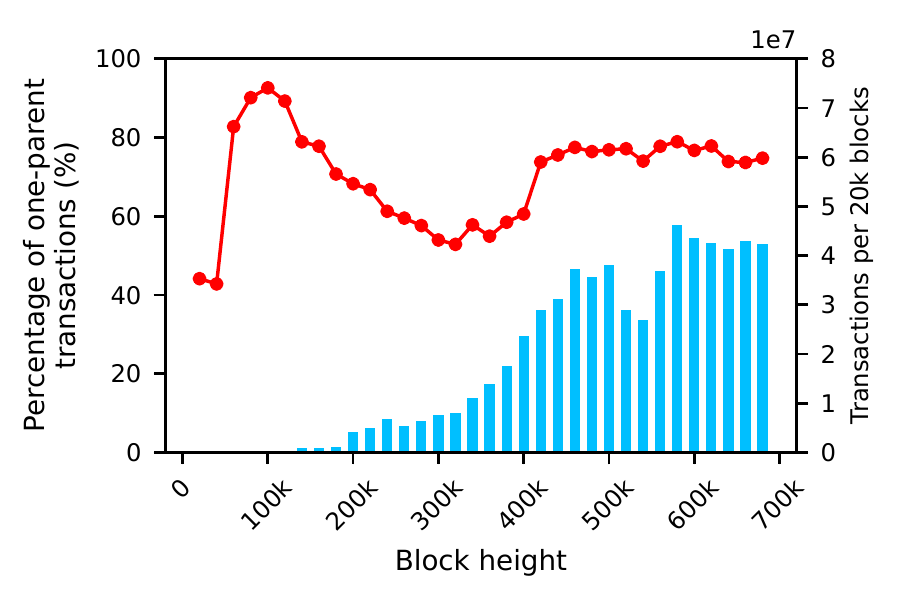}}
  \caption{The percentage of transactions with one parent}
   \label{fig:one_parent_tx_percent}
\end{figure}

Fig. \ref{fig:input_tx_cnt} shows that parent transaction counts conform to a power-law distribution, which means that transactions with a few parents occur more often than transactions with many parents. We are particularly interested in transactions with one parent because a placement algorithm can easily make such transactions single-shard. 
By analyzing block 0$\sim$680k (mined in April 2021 \cite{block680k}), we found that one-parent transactions account for a significant proportion throughout the history of Bitcoin and remain approximately 75\% since block 420k as shown in Fig. \ref{fig:one_parent_tx_percent}.
The percentage changes a lot at heights lower than 100k because blocks include only a few transactions at the start of Bitcoin as shown by the blue bars in Fig \ref{fig:one_parent_tx_percent}.

\subsection{Cost of Cross-shard Transactions} \label{sec:cost_aly}
In this section, we compare the execution time of single-shard transactions and cross-shard transactions by quantitative measurements. 
An OmniLedger-like sharding protocol is implemented based on Bitcoin Core \cite{bitcoinCore}, and 27k Bitcoin transactions (from block 601000 to block 601009) are replayed in a 2-shard environment. More details about the implementation and the testbed will be given in Section \ref{sec:eval}. For a single-shard transaction, the client sends a \texttt{TX} request containing the transaction directly to the output shard. For a cross-shard transaction, the client sends a \texttt{LOCK} request to every input shard and a \texttt{COMMIT} request to the output shard, provided all input shards reply with positive locking results (see Fig. \ref{fig:atomix}). 
The measurement results show that the median processing time for a \texttt{TX} request, a \texttt{LOCK} request, and a \texttt{COMMIT} request are 211 $\mu$s, 438 $\mu$s, and 259 $\mu$s, respectively. 

\begin{figure}[htbp]
\centering
\subfloat[TX, LOCK, and COMMIT]{\includegraphics[trim={0 0cm 0cm 0.8cm}, width=0.48\linewidth]{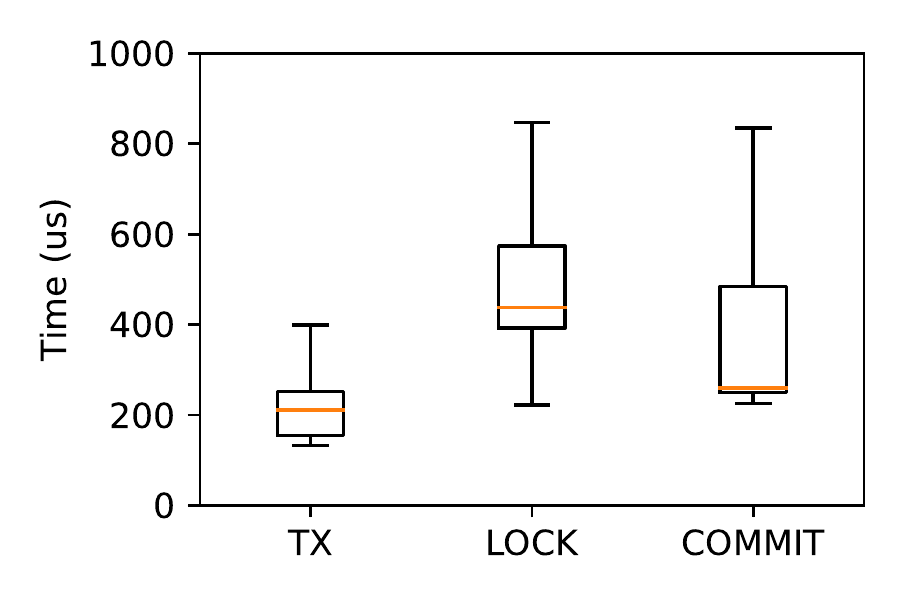}%
\label{fig:overall_time}}
\hfill
\subfloat[Breakdown of TX]{\includegraphics[trim={0 0cm 0cm 0.8cm}, width=0.48\linewidth]{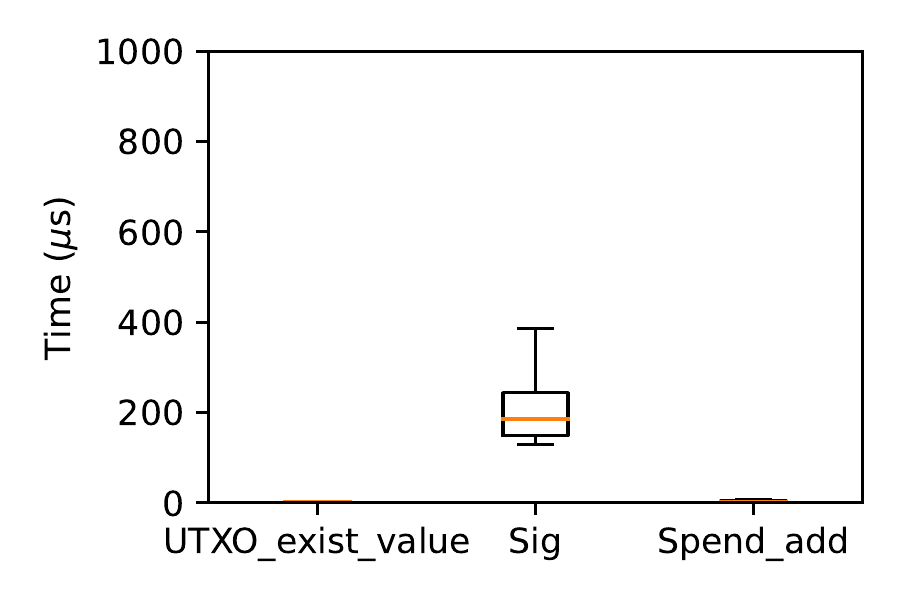}%
\label{fig:tx_detail}} \\
\subfloat[Breakdown of LOCK]{\includegraphics[trim={0 0cm 0cm 0cm}, width=0.48\linewidth]{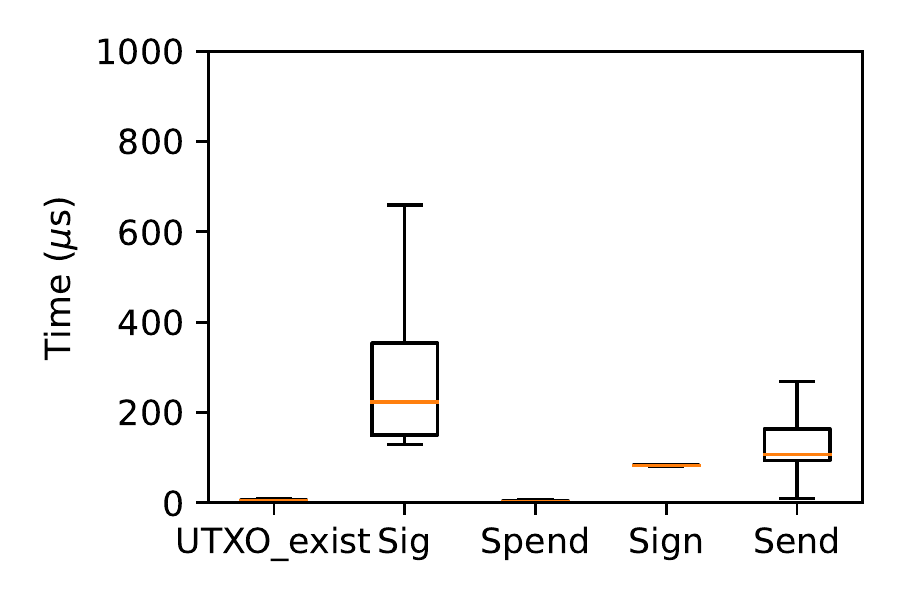}%
\label{fig:lock_detail}} 
\hfill
\subfloat[Breakdown of COMMIT]{\includegraphics[trim={0 0cm 0cm 0cm}, width=0.48\linewidth]{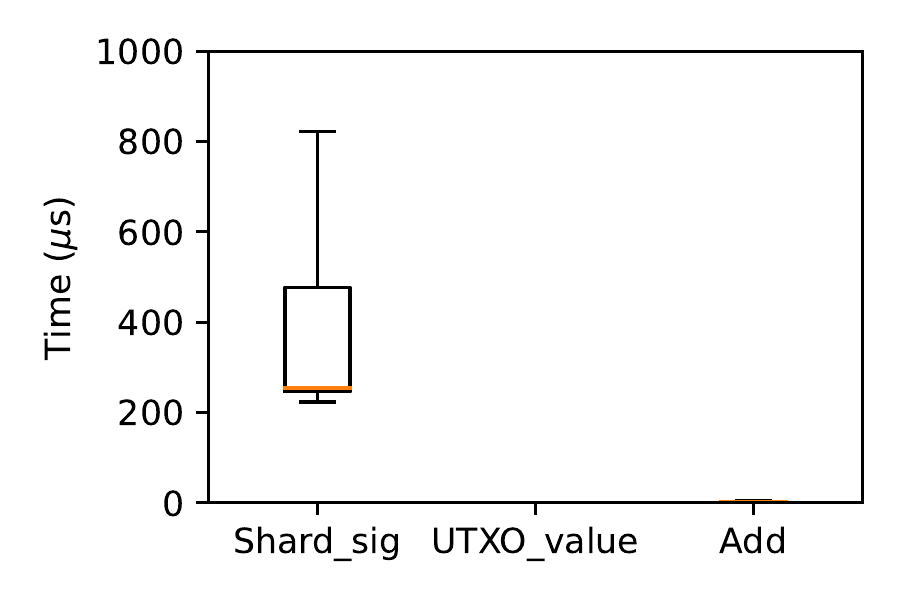} %
\label{fig:commit_detail}}
\caption{Request processing time (2 shards)}
\label{fig:time_measure}
\end{figure}

To understand why \texttt{LOCK} and \texttt{COMMIT} requests take a longer time to process than \texttt{TX} requests, we further measure the processing time of each step in request execution.
In an UTXO-based blockchain, a node checks for three conditions when verifying a transaction: 1) the input UTXOs exist and are unspent, 2) the total input value is not less than the total output value, and 3) the transaction includes the correct signatures of the input UTXO owners. 
Provided the transaction meets all three conditions, the node will execute it by removing the input UTXOs and add the output UTXOs to the system state.  
In Fig. \ref{fig:tx_detail}, the \emph{UTXO\_exist\_value} label  corresponds to the first two transaction verification steps, the \emph{Sig} label to the third verification step, and the \emph{Spend\_add} label to the system state update.  Fig. \ref{fig:tx_detail} shows that checking signatures of input UTXOs dominates the processing time of a single-shard transaction.
In Fig. \ref{fig:lock_detail}, the \emph{UTXO\_exist} label corresponds to the first transaction verification step. Note that \texttt{LOCK} request processing does not include value checking because an input shard is not aware of the value of input UTXOs in other input shards. Thus, value checking is done by  the output shard during \texttt{COMMIT} request processing. Similarly, system state update is separated into two parts: removing input UTXOs in \texttt{LOCK} request processing, and adding the output UTXOs in \texttt{COMMIT} request processing. Compared with \texttt{TX} request processing, \texttt{LOCK} request processing includes two additional steps: signing the lock result (labeled as \emph{Sign}), and sending the result to the client (labeled as \emph{Send}). These two steps are comparable in processing time to input UTXO signature checking. Lastly, \texttt{COMMIT} request processing is dominated by verifying the signatures of lock results, which are produced by input shards. The step is labeled as \emph{Shard\_sig} in Fig. \ref{fig:commit_detail}.

As a result, if a one-parent transaction becomes cross-shard as a result of being placed to a different shard than its parent, its processing time would be more than tripled due to input shard authentication and communication overhead.


\section{Case Study: OptChain}
OptChain is a client-side blockchain transaction placement algorithm.
A client runs the algorithm to compute the output shard ID before sending its transaction to servers. 
OptChain has been proven to have better performance than Metis, an offline graph partitioning algorithm, and a greedy algorithm that places transactions to the shard with most of its parents.
\subsection{Overview of OptChain}\label{sec:overview}
\begin{figure}[htbp]
  \centerline{\includegraphics[trim={0 0cm 0cm 0cm}, width=0.6\linewidth]{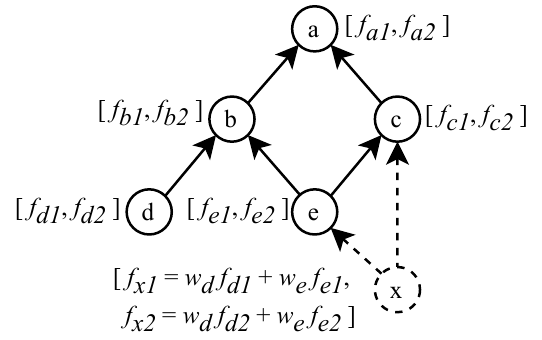}}
  \caption{Principle of OptChain. The fitness-score array of $x$ is a weighted sum of the fitness-score arrays of $x$'s parents.}
   \label{fig:optchain_dia}
\end{figure}

We summarize the design principles of OptChain in this section. The goal of OptChain is to reduce cross-shard transactions while maintaining load balance. 
To reduce cross-shard transactions, OptChain builds a transaction graph with transactions as vertices and transaction dependencies as edges.
Therefore, from the perspective of graph theory, the transaction placement problem is a streaming graph partitioning problem \cite{stanton2012streaming}. 
OptChain associates each transaction with a fitness-score array of size $n_s$, which is the number of shards in the system. Each element in the array reflects the fitness between the transaction and the corresponding shard. 
Based on PageRank \cite{page1999pagerank}, OptChain computes a child transaction's fitness-score array as a weighted sum of its parents' fitness-score arrays. 
For example, in Fig. \ref{fig:optchain_dia}, when a new transaction $x$ arrives in a two-shard system, OptChain calculates the fitness score of the first shard as $f_{x1} = w_d f_{d1} + w_e f_{e1}$, since transaction $d$ and $e$ are the two parents of $x$. The weight of the a parent is $\nicefrac{1}{n_c}$, where $n_c$ is the number of child transactions of this parent; so, in the example, the weights of both parent transactions, namely $w_d$ and $w_e$, equal one.
The fitness score of the second shard $f_{x2}$ is calculated the same way.

To account for load balance, OptChain divides a fitness score by the corresponding partition size (i.e., the number of transactions placed to the shard), and 
calls the resultant value the Transaction-to-Shard (T2S) score of the shard. In addition, OptChain requires clients to frequently sample servers for communication latency and server-side transaction queue length in order to balance loads dynamically. We refer to this design as feedback load balancing because it resembles the idea of feedback control in control theory---use the measured shard latency to adjust how many transactions should be placed to the shard. The T2S score and feedback load balancing jointly determine the output shard of a given transaction.

\subsection{Partitioning Quality}\label{sec:part_quality}
We assess partitioning quality using two metrics: cross-shard transaction ratio, and load imbalance, which is defined as the maximum partition size minus the minimum partition size.
To study whether OptChain can produce high-quality partitions under various workloads, we employ four transaction sets that are detailed in Table \ref{tab:dataset} to evaluate \emph{OptChain-T2S}, which is Optchain without feedback load balancing. OptChain-T2S is used instead of OptChain because the feedback load balancing feature demands running servers, whereas OptChain-T2S is independent of server states, enabling us to obtain the partitioning quality without running servers. 
This independence is particularly convenient when evaluating the partition quality under a large $n_s$. Besides, OptChain-T2S is the core of OptChain; therefore, understanding the performance of OptChain-T2S is definitely helpful in understanding the performance of OptChain.
We will show the impact of feedback load balancing in Section \ref{sec:eval}. 
Placement results for transactions prior to the start block height of a dataset are generated using OptChain-T2S beforehand.

\begin{table}
\centering
  \caption{Four datasets}
  \label{tab:dataset}
  \begin{tabular}{l|l|l}
    \hline
    Name & Block heights & Transaction count$^{\mathrm{a}}$\\
    \hline
    dataset 1 & [0, 200k)  & 7,316,308 \\
    \hline
    dataset 2 & [200k, 227k)  & 7,371,053 \\
    \hline
    dataset 3 & [227k, 252k)  & 7,316,337 \\
    \hline
    dataset 4 & [252k, 275k)  & 7,238,332 \\
    \hline
    \multicolumn{3}{l}{$^{\mathrm{a}}$Coinbase trancations are excluded.}
\end{tabular}
\end{table}

\begin{figure}[htbp]
\centering
\subfloat[dataset 1]{\includegraphics[trim={0.2cm 0.4cm 0cm 0.5cm}, width=0.5\linewidth]{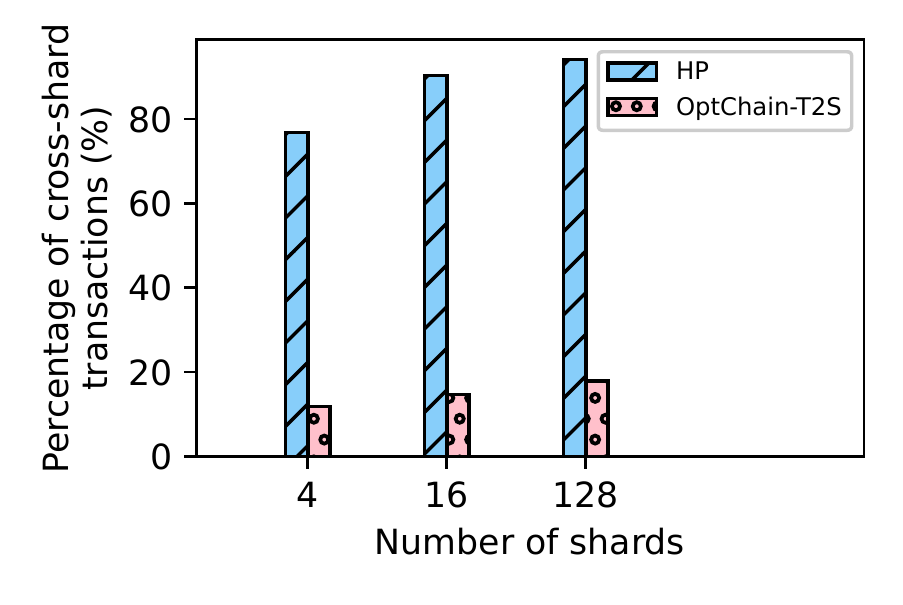}%
\label{fig:cross_shard_cnt1}}
\subfloat[dataset 2]{\includegraphics[trim={0 0.4cm 0.2cm 0.5cm}, width=0.5\linewidth]{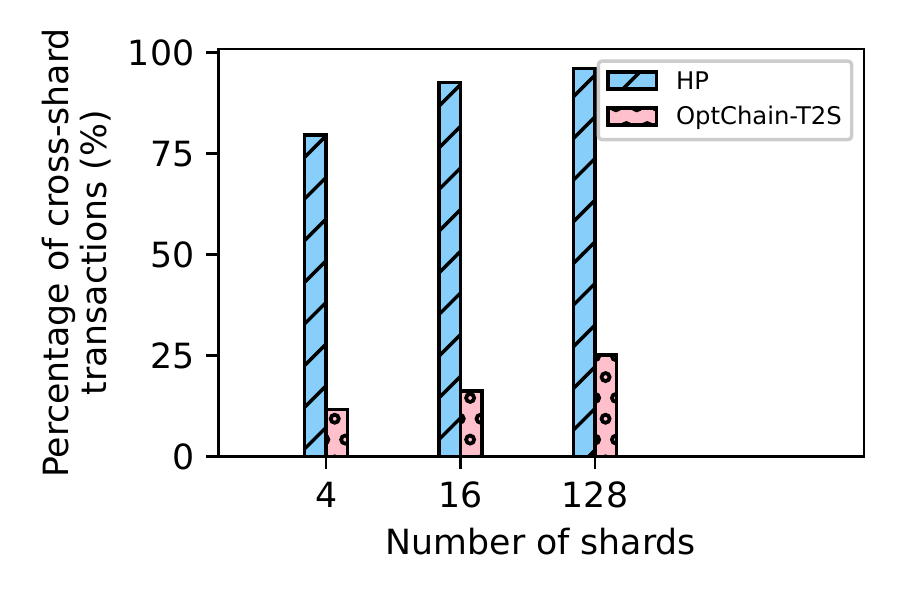}%
\label{fig:cross_shard_cnt2}} \\
\subfloat[dataset 3]{\includegraphics[trim={0.2cm 0.4cm 0cm 0.4cm}, width=0.5\linewidth]{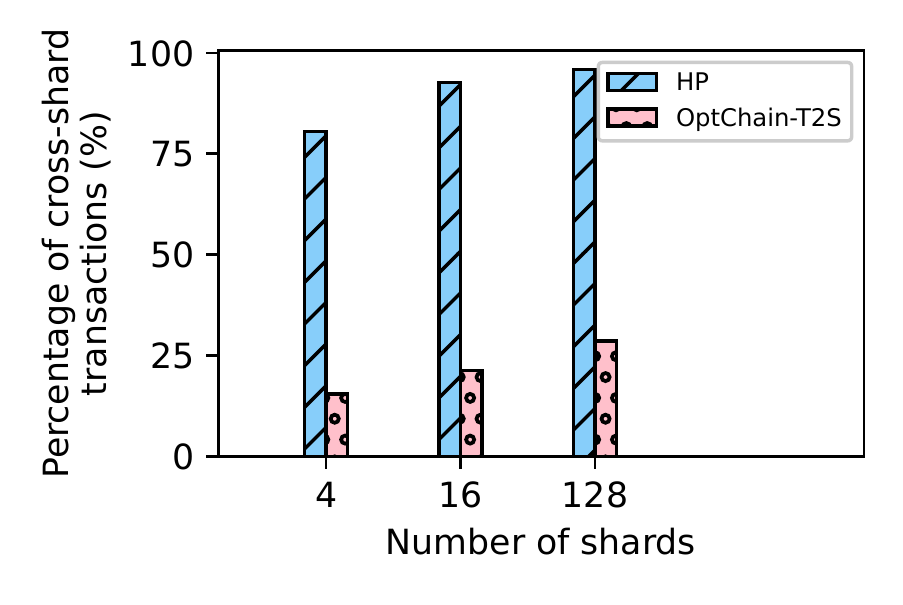}%
\label{fig:cross_shard_cnt3}}
\subfloat[dataset 4]{\includegraphics[trim={0 0.4cm 0.2cm 0.4cm}, width=0.5\linewidth]{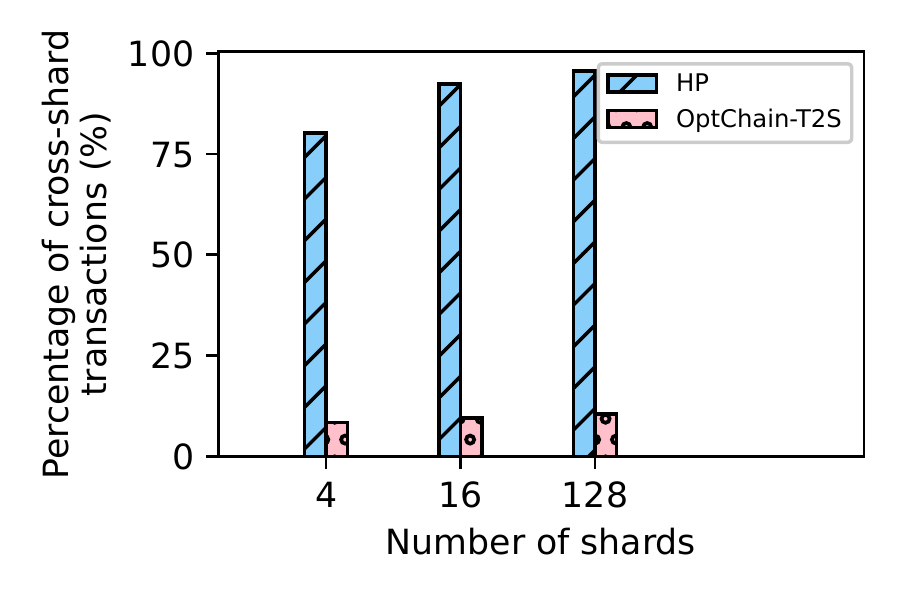}%
\label{fig:cross_shard_cnt4}}
\caption{Cross-shard transaction ratio}
\label{fig:cross_shard_ratio}
\end{figure}

\begin{figure}[htbp]
\centering
\subfloat[dataset 1]{\includegraphics[trim={0.2cm 0.4cm 0cm 0.8cm}, width=0.5\linewidth]{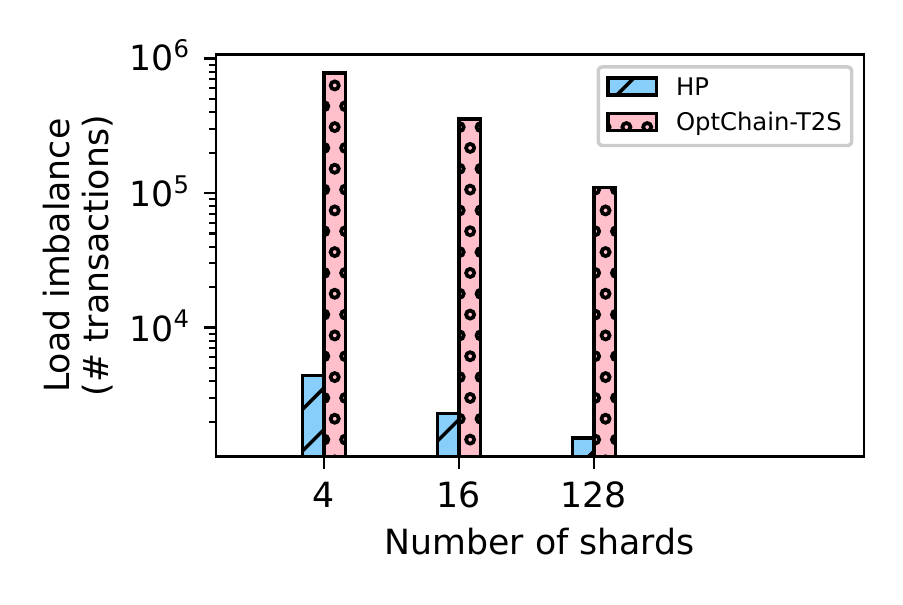}%
\label{fig:imbalance1}}
\subfloat[dataset 2]{\includegraphics[trim={0cm 0.4cm 0.2cm 0.8cm}, width=0.5\linewidth]{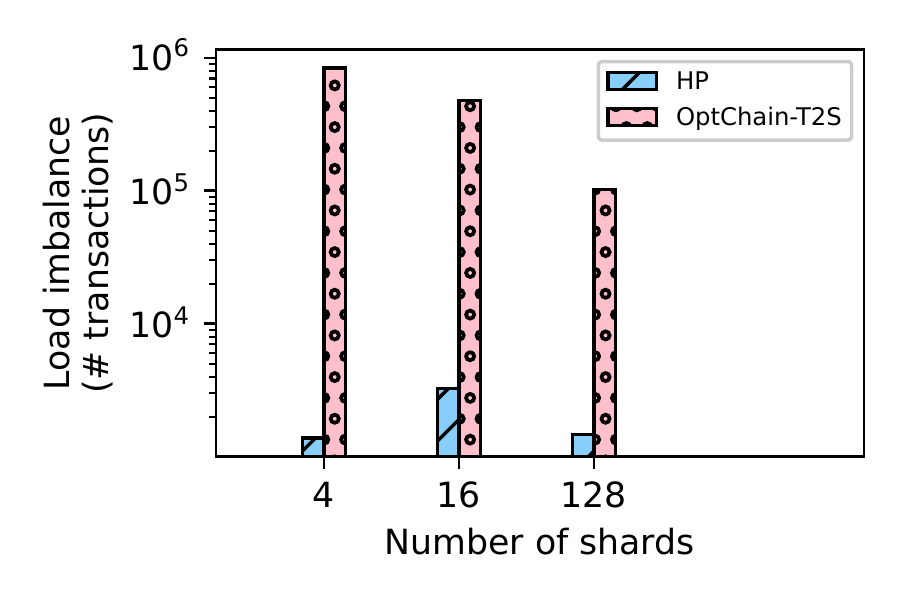}%
\label{fig:imbalance2}} \\
\subfloat[dataset 3]{\includegraphics[trim={0.2cm 0.4cm 0cm 0.4cm}, width=0.5\linewidth]{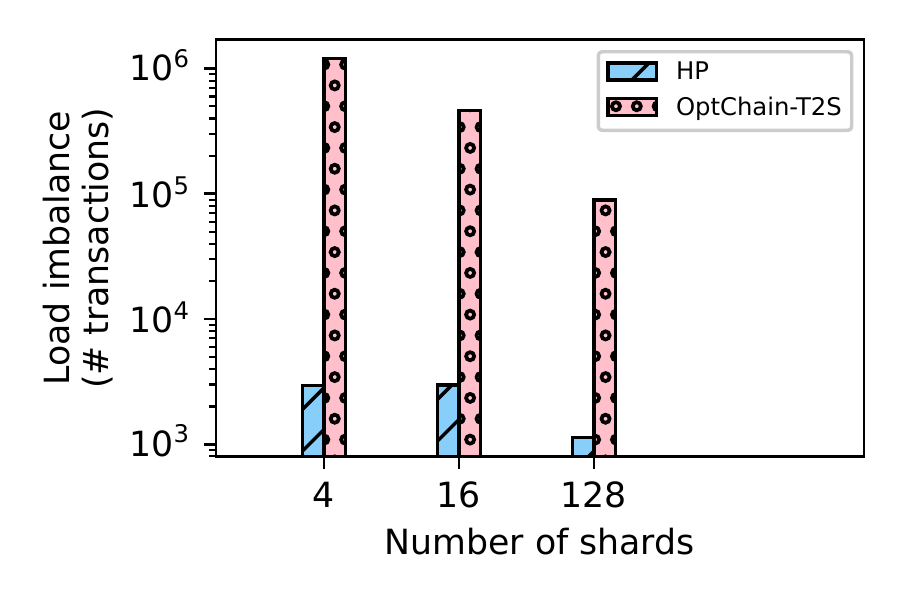}%
\label{fig:imbalance3}}
\subfloat[dataset 4]{\includegraphics[trim={0cm 0.4cm 0.2cm 0.4cm}, width=0.5\linewidth]{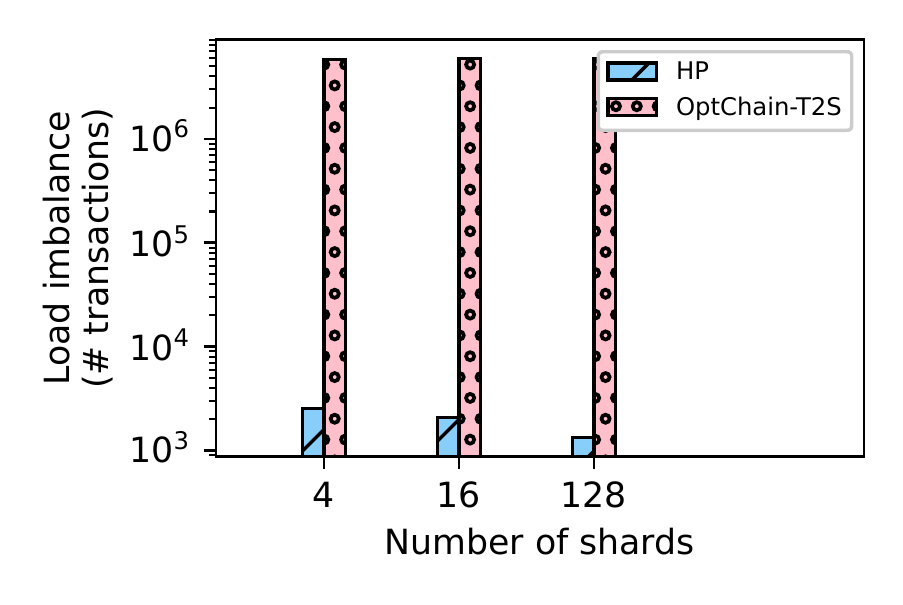}%
\label{fig:imbalance4}}
\caption{Load imbalance}
\label{fig:imbalance}
\end{figure}

Fig.~\ref{fig:cross_shard_ratio} and Fig.~\ref{fig:imbalance} compare the cross-shard transaction ratio and load imbalance of OptChain-T2S and HP.
For all datasets, OptChain-T2S produces significantly fewer cross-shard transactions than HP. 
From Fig. \ref{fig:imbalance}, one can see that HP does extremely well at load balancing. Although Optchain-T2S does not evenly distribute transactions to shards, the load imbalance decreases as the number of shards increases. However, in Fig. \ref{fig:imbalance4}, some shard still receives considerably more transactions than others even with 128 shards. 
Considering the fact that OptChain-T2S produces noticeably less cross-shard transactions for dataset 4 than for the other three datasets as shown in Fig \ref{fig:cross_shard_ratio}, we suspect that dataset 4 may have some special traits that break Optchain-T2S's load balancing mechanism and cause the vast majority of transactions to be placed to the same shard.

\begin{figure}[htbp]
\centering
\subfloat[HP]{\includegraphics[trim={0 0.2cm 0cm 0.3cm}, width=0.8\linewidth]{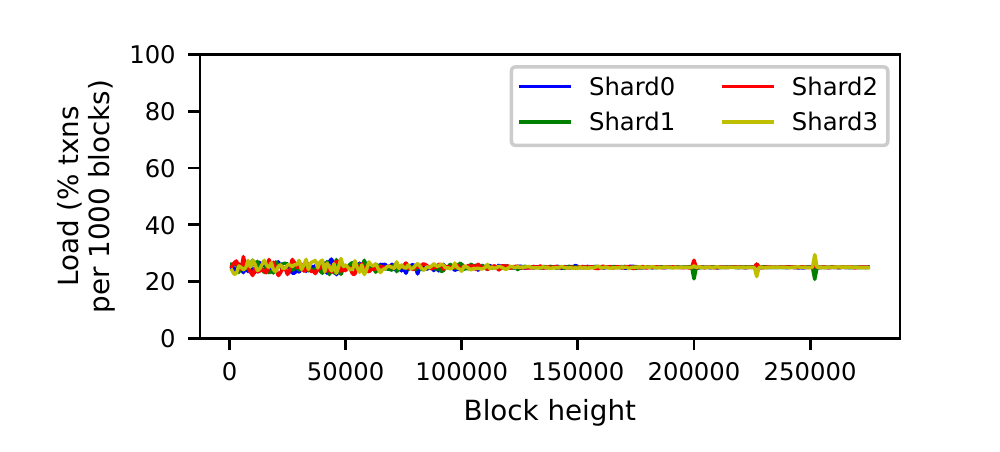}%
\label{fig:load_hp}}

\subfloat[OptChain-T2S]{\includegraphics[trim={0 0.2cm 0cm 0.3cm}, width=0.8\linewidth]{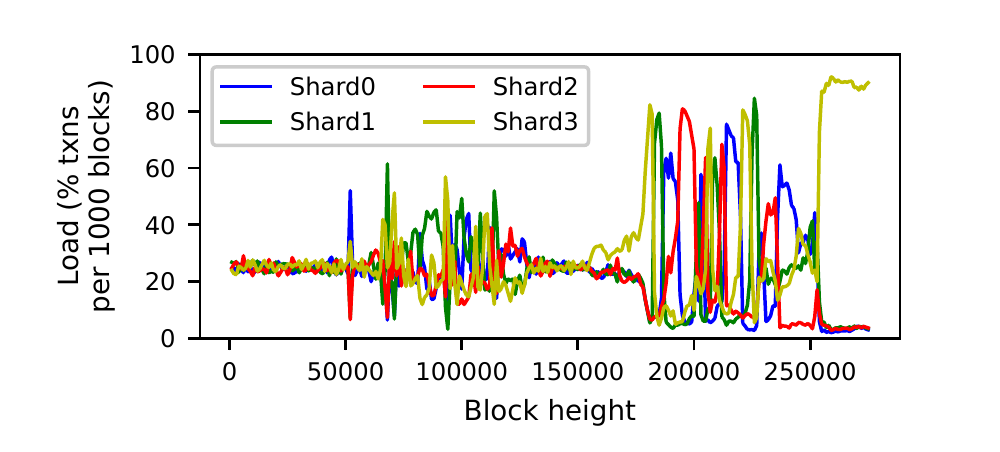}%
\label{fig:load_optchain_t2s}}

\caption{Dynamic shard loads (4 shards)}
\label{fig:dyn_load}
\end{figure}

To confirm the above hypothesis, we analyze the dynamic shard loads in a 4-shard environment as illustrated in Fig. \ref{fig:dyn_load}, which implies that OptChain-T2S fails to balance loads since block 253k. Unsurprisingly, HP balances loads remarkably well in that every shard constantly receives about 25\% of the transactions. In Fig. \ref{fig:load_hp}, the three small spikes between block 200k and 260k are the starting points of datasets 2$\sim$4, and the glitches at the beginning result from small block sizes. 
Another observation is that, in Fig. \ref{fig:load_optchain_t2s}, all curves are quite flat in range of block [0, 50k) and block [125k, 175k), but fluctuate a lot in the range of [70k, 120k) and [180, 250k). We find that this pattern occurs with transactions spending the output UTXOs of their immediate predecessors, which are ordered right before them in the same block. OptChain-T2S tends to place such transactions to the same shard as their predecessors. 
As a result, a sequence of such transactions causes a shard to receive more transactions than other shards temporarily.
For example, starting from the 326th transaction in block 177253,  
each of the 267 subsequent transactions is a child of its immediate predecessors \cite{spendLastTx}. 
Table \ref{tab:spending_pred} shows that such transactions account for a relatively high percentage whenever the shard load curves fluctuate drastically.

\begin{table}[htbp]
\centering
  \caption{Percentage of Transactions Consuming the UTXO Produced by their Immediate Predecessors}
  \label{tab:spend_last}
  \begin{tabular}{l|>{\centering\arraybackslash} m{3.7cm}}
    \hline
    Block height & Transactions depending on their immediate predecessors \\
    \hline
    [0, 50k)  &  0.1\% \\
    \hline
    [70k, 120k)  & 18.6\% \\
    \hline
    [125k, 175k)  & 6.1\% \\
    \hline
    [180k, 250k)  & 21.3\% \\
    \hline
    [253k, 275k)  & 14.6\% \\
    \hline
\end{tabular}
\label{tab:spending_pred}
\end{table}

\begin{figure}[htbp]
  \centerline{\includegraphics[trim={0 0.2cm 0cm 0cm}, width=0.8\linewidth]{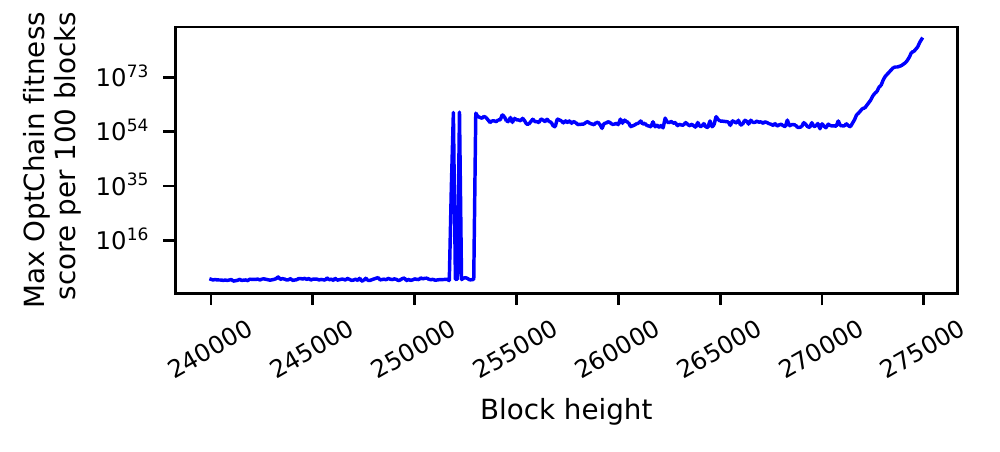}}
  \caption{Variation in OptChain fitness score}
   \label{fig:max_fit_score}
\end{figure}

\begin{figure}[htbp]
\centering
\subfloat[$x$ amplifies fitness scores.]{\includegraphics[trim={0 0 0 0.3cm}, width=0.49\linewidth]{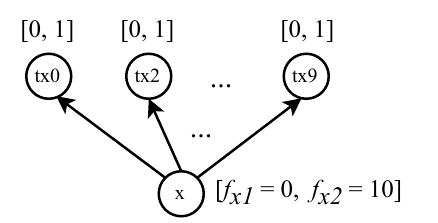}%
\label{fig:agg_tx1}}
\hfill
\subfloat[$x$ dominates $y$'s placement result.  ]{\includegraphics[trim={-0.12cm 0 -0.12cm 0.3cm}, width=0.5\linewidth]{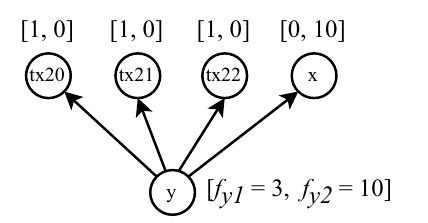}%
\label{fig:agg_tx2}}
\caption{An example of an aggregating transaction}
\label{fig:agg_tx}
\end{figure}

\begin{figure}[htbp]
  \centerline{\includegraphics[trim={0 0.2cm 0cm 0cm}, width=0.8\linewidth]{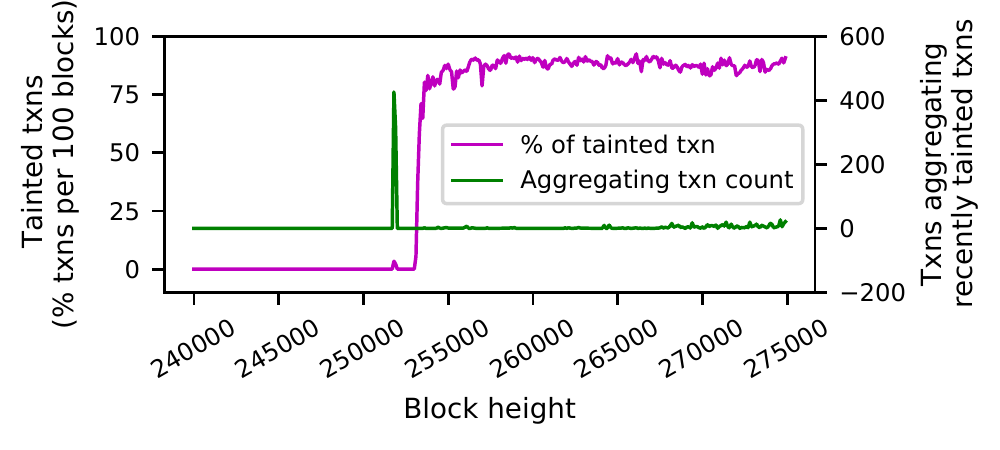}}
  \caption{Tainted transactions and aggregating transactions}
   \label{fig:tainted_and_aggregate}
\end{figure}

To further understand why OptChain-T2S suddenly loses its load balancing capability, we analyze transaction fitness scores and find that some transactions have abnormally high fitness scores since block 251750 (see Fig. \ref{fig:max_fit_score}). These high fitness scores are created by transactions that ``aggregate" many previous transactions placed to the same shard. Fig. \ref{fig:agg_tx} illustrates a simplified scenario of such, where transaction $x$ has 10 parents whose fitness-score array are all [0, 1] (i.e., the system has two shards, and the second shard has the higher fitness score for all the parents). $x$'s fitness-score array can be computed as [0, 10] by following the rules described in Section \ref{sec:overview}. Note that the second element of $x$'s fitness-score array is 10 times higher than that of its parents. In Fig. \ref{fig:agg_tx2}, transaction $y$ is a child transaction of $x$, and its fitness-score array is [3, 10]. 
Provided the two shards have the same load, $y$ will be placed to the second shard even though three out of four of its parents are in the first shard. Moreover, $y$ gets ``tainted" by $x$ (i.e., $y$ inherits the high fitness score of the second shard from $x$) and will propagate the high fitness score to its children.
Between block 251750 and block 251840, many transactions aggregate recently tainted transactions (Fig. \ref{fig:tainted_and_aggregate}) and collectively amplify the fitness score by 50 orders of magnitude. 
In other words, aggregating transactions amplify fitness scores, and the resultant high fitness scores propagate to other transactions through dependencies. 
By comparing Fig. \ref{fig:tainted_and_aggregate} with Fig. \ref{fig:max_fit_score}, one can see that 1) whenever there are tainted transactions, the fitness score in Fig. \ref{fig:max_fit_score} is at a high level, and 2) whenever there are transactions that aggregate recently tainted transactions, the fitness score grows. 

OptChain-T2s's failure to handle dataset 4 proves that PageRank is not completely suitable for the transaction placement problem.
PageRank is originally designed by Google to identify important websites. A website referenced by a lot of other websites is deemed important. However, this notion does not apply to the transaction graph. A transaction with many parent transactions should not be important and dominates the placement result of its child transactions. Besides, in PageRank, a website is also deemed  important if it is referenced by an important website. When mapped from a website graph to a transaction graph, this rule becomes ``a transaction is important if one of its parent transactions is important". 
Based on this rule, a transaction inherits high fitness scores from its aggregating parents and heavily ``attract"s its descendants to its shard.

\begin{figure}[htbp]
  \centerline{\includegraphics[trim={0 0.2cm 0cm 0cm}, width=0.8\linewidth]{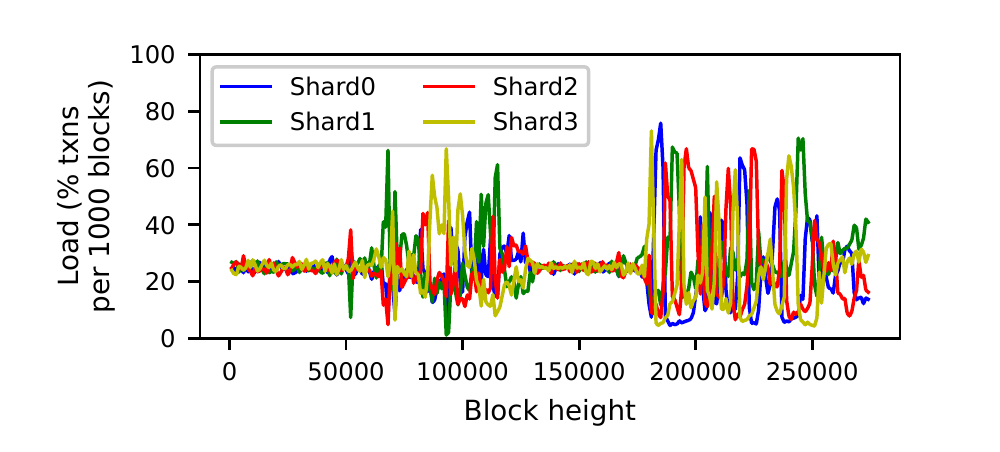}}
  \caption{Dynamic shard loads of OptNorm (4 shards)}
   \label{fig:load_optnorm}
\end{figure}

Knowing the shortcoming of OptChain-T2S, we propose a quick fix---\emph{OptNorm}, which normalizes fitness scores. Specifically, given a transaction, OptNorm computes the fitness score array and selects the output shard in the same way as OptChain-T2S does.
Next, however, OptNorm normalizes the fitness-score array, i.e., dividing each element by the sum of all elements. The normalization guarantees future transactions will have parent transactions of equal importance since every parent transaction's fitness score array has an element-wise sum of one. Fig. \ref{fig:load_optnorm} illustrates that OptNorm balance loads better than OptChain-T2S and can survive aggregating transactions.

\section{Evaluation}\label{sec:eval}
We implemented OptChain as well as an OmniLedger-like sharding protocol based on Bitcoin Core \cite{bitcoinCore}.
We made the same change to the OmniLedger protocol as the OptChain paper does to overcome the bandwidth limit---clients send transactions directly to the input and output shards instead of gossiping the transaction.
Experiments are done on a local cluster. 64 VMs (each with 2 CPUs and 12GB memory) are created on 16 machines, each of which has dual Xeon E5-2620 at 2.1 GHz (12 cores) and 64GB RAM. Each VM runs an instance of our modified Bitcoin Core and is used as a blockchain node. 
One client runs on another machine with dual Xeon E5-2630 at 2.6 GHz (12 cores, 2 hyperthreads per core) and 256GB RAM.
The client reads historical transactions in Bitcoin block [250k, 260k) with 2.9 million transactions in total, and sends them to the peers at a rate of 4.5k tps. Transactions in this range are used in order to evaluate the impact of aggregating transactions on system performance.   
A latency of 100 ms is imposed on every link between peers, and the bandwidth is limited to 500 Mbps.
We use this relatively high bandwidth because we notice that OptChain-T2S places most transactions in one shard, and low network bandwidth will be a bottleneck for that shard.

\begin{figure}[htbp]
\centering
\subfloat[Throughput]{\includegraphics[trim={0.2cm 0.3cm 0.1cm 0.6cm}, width=0.5\linewidth]{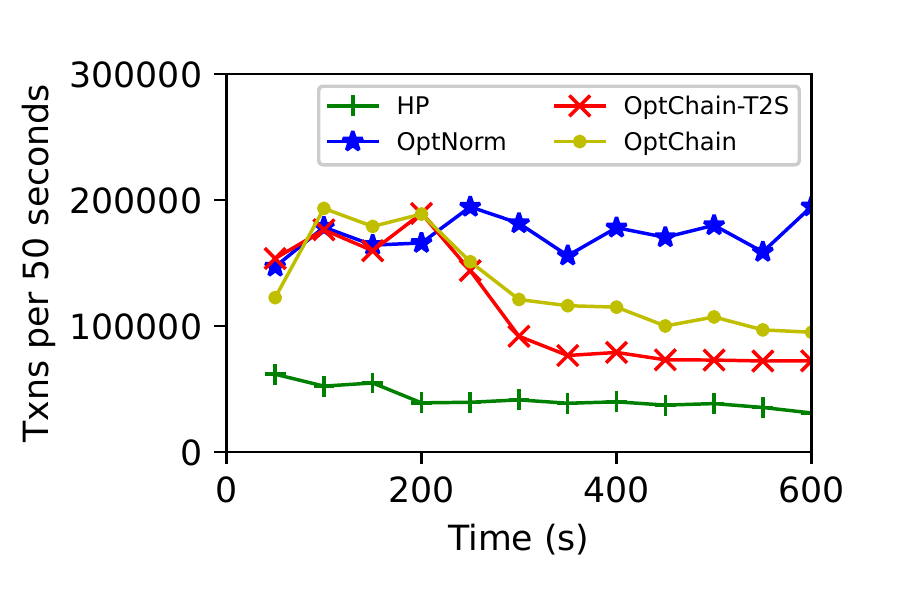}%
\label{fig:thru_250k}}
\subfloat[OptNorm]{\includegraphics[trim={0.2cm 0.3cm 0cm 0.6cm}, width=0.5\linewidth]{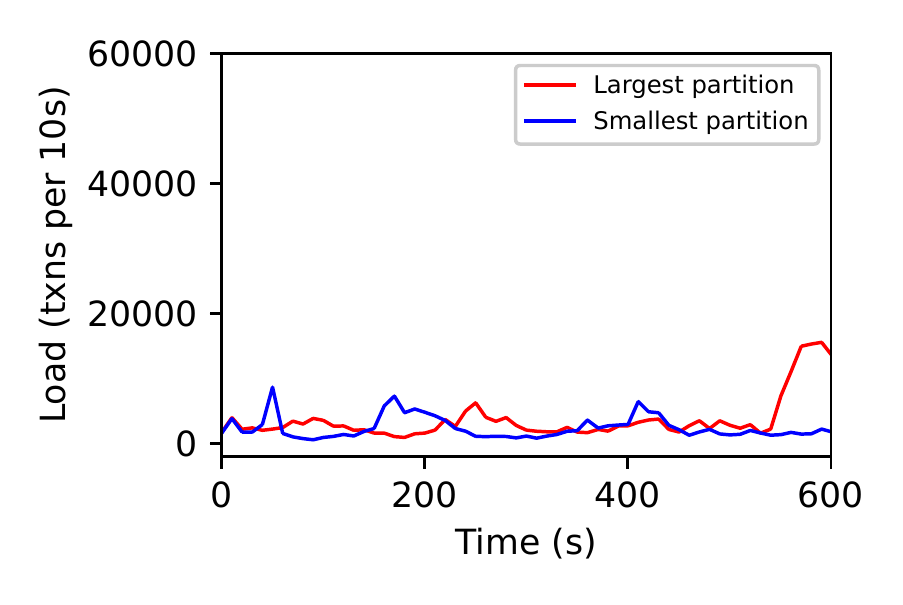}%
\label{fig:load_two_lines_optnorm}}\\
\subfloat[OptChain-T2S]{\includegraphics[trim={0 0.3cm 0.2cm 0.3cm}, width=0.5\linewidth]{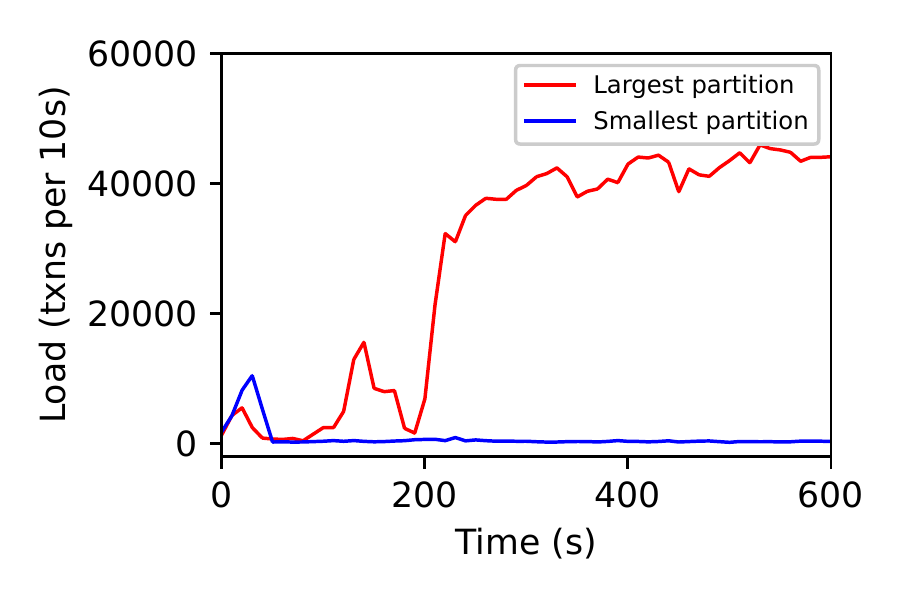}%
\label{fig:load_two_lines_optchainT2S}} 
\subfloat[OptChain]{\includegraphics[trim={0 0.3cm 0.2cm 0.3cm}, width=0.5\linewidth]{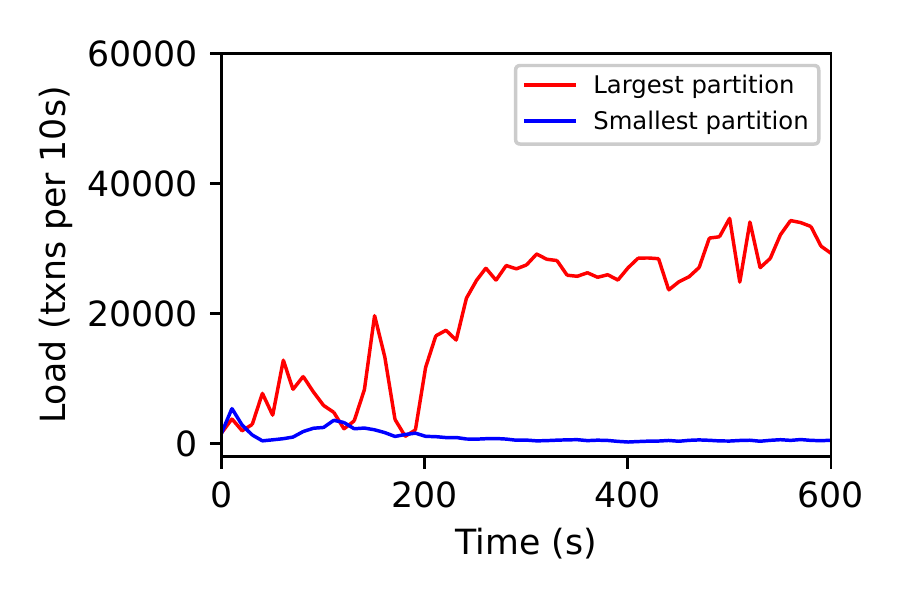}%
\label{fig:load_two_lines_optchainLB}}
\caption{Throughput and loads comparisons (16 shards)}
\label{fig:perf_250k}
\end{figure}

Fig. \ref{fig:thru_250k} demonstrates that feedback load balancing mitigates the performance degradation of OptChain-T2S but cannot eliminate the degradation. 
By comparing Fig. \ref{fig:load_two_lines_optchainT2S} and \ref{fig:load_two_lines_optchainLB}, one can see that, although the load of the largest partition under OptChain is lower than that under OptChain-T2S, it still surges at around 200s, exactly when the throughput starts to drop. In contrast, OptNorm distributes load fairly evenly at 200s, which confirms that it can effectively overcome the aggregating transaction issue of OptChain.
The overall throughput of HP and OptNorm are 865 tps and 3487 tps, respectively, which means OptNorm improves the system performance by 4x.

\begin{figure}[htbp]
\centering
\subfloat[Throughput]{\includegraphics[trim={0.2cm 0.3cm 0.1cm 0.6cm}, width=0.5\linewidth]{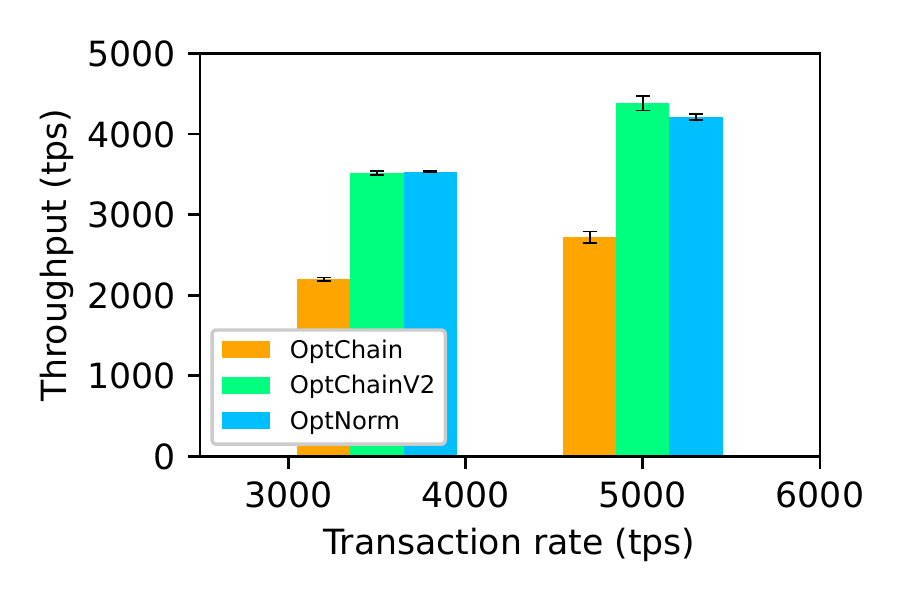}%
\label{fig:thru_optV2}}
\subfloat[Latency (3.5k-tps transaction rate)]{\includegraphics[trim={0.2cm 0.3cm 0cm 0.6cm}, width=0.5\linewidth]{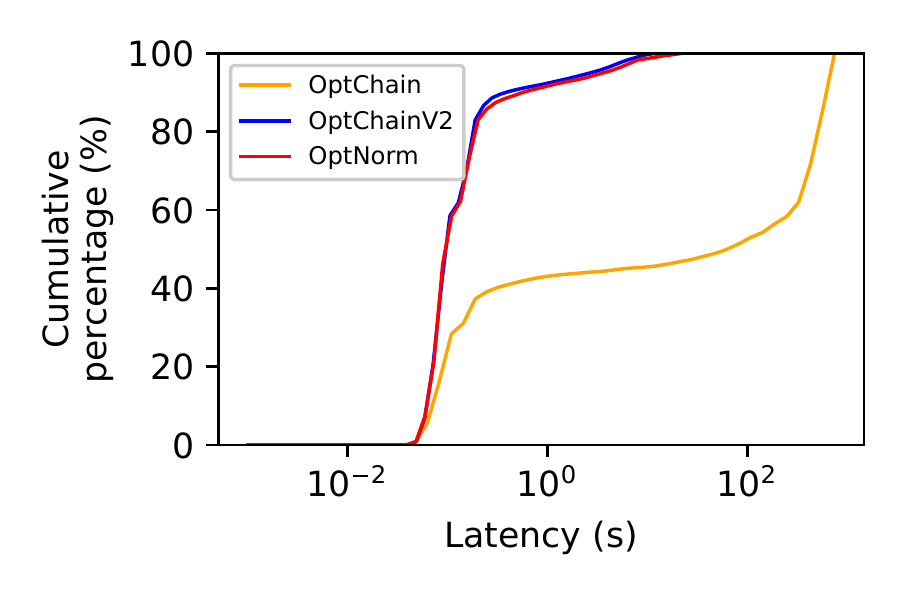}%
\label{fig:latency_optV2}}
\caption{Performance comparison (16 shards)}
\label{fig:perf_optV2}
\end{figure}

The authors of OptChain noticed our research and created a new version of the OptChain paper to update their algorithm description (referred to as \emph{OptChainV2}) \cite{optchain-v2}. Specifically, OptChainV2 allows multiple edges between two vertices in the transaction graph, and changes the definition of $N_{in}$ ($N_{out}$) from a set of input (output) transactions to a multiset of input (output) transactions.  
For completeness, this paper also evaluates the performance of OptChainV2 using Bitcoin block [250, 260) and compares the result with OptNorm and OptChain. 
To measure latency, all transactions must be completed within a reasonable amount of time, so we reduce link delay to 10ms and warm up  database cache before each test. This testing environment modification equally benefits all placement algorithms. 
Fig. \ref{fig:perf_optV2} shows, when the client sends transactions at a rate of 3.5k tps, both OptChainV2 and OptNorm can achieve a 3.5k-tps throughput, whereas OptChain can reach only 2.2k tps. Also, the 50th percentile latency of OptChainV2 and OptNorm are almost the same, both 99.9\% shorter than the 50th percentile latency of OptChain. On the other hand, when the system is overloaded with a 5k-tps transaction rate, OptChainV2 achieves  4\% higher throughput  than OptNorm. 

\section{Conclusion}
In conclusion,
cross-shard transactions are expensive due to input shard authentication and communication overhead. 
Placement algorithms considering transaction dependencies  can significantly reduce cross-shard transaction ratio and thus boost system performance. 
The original OptChain does not balance loads well under workloads that include many aggregating transactions and their descendants. The proposed OptNorm is a simple yet effective fix. OptChainV2 does not exhibit the shortcoming.

\balance
\bibliographystyle{IEEEtran}
\bibliography{reference}
\end{document}